\documentclass[conference]{IEEEtran}
\IEEEoverridecommandlockouts
\usepackage{latexsym}
\usepackage{graphicx}
\usepackage{amsfonts,amssymb,amsmath,bm}
\usepackage{geometry}
\def\BibTeX{{\rm B\kern-.05em{\sc i\kern-.025em b}\kern-.08em T\kern-.1667em\lower.7ex\hbox{E}\kern-.125emX}}

\usepackage{cite}
\usepackage{amsmath,amssymb,amsfonts}
\usepackage{algorithmic}
\usepackage{graphicx}
\usepackage{textcomp}
\usepackage{xcolor}
\usepackage{float}
\usepackage{multicol}
\usepackage{verbatim}
\usepackage{amsmath,amssymb}
\usepackage{algorithm}
\usepackage{float} 
\usepackage{booktabs} 
\usepackage{tabularx} 
\usepackage{caption}  
\usepackage{graphicx} 
\usepackage{subcaption} 

\usepackage[acronym]{glossaries}
\newcommand{\newac}{\newacronym}

\makeglossaries

\newac{los}{\text{LoS}}{line-of-sight}
\newac{nlos}{\text{NLoS}}{non-line-of-sight}

\DeclareMathAlphabet{\mathbit}{OML}{cmr}{bx}{it}
\DeclareMathAlphabet{\mathsf}{OT1}{cmss}{m}{n}
\DeclareMathAlphabet{\mathTXf}{OT1}{cmss}{bx}{it}

\DeclareMathOperator{\Transpose}{T}

\DeclareMathOperator*{\argmax}{arg\ max}

\DeclareMathAlphabet{\mathpzc}{OT1}{pzc}{m}{it}
\newcommand{\TA}{{\text{TA}}}

\newcommand{\Tr}{{\Transpose}}

\begin{document}

\title{First Results on UAV-aided User Localization Using ToA and OpenAirInterface in 5G NR
}

\author{
\IEEEauthorblockN{Omid Esrafilian$^{(1)\,*}$, Rakesh Mundlamuri $^{(1)\,*}$, Florian Kaltenberger$^{(1) (2)}$, Raymond Knopp$^{(1)}$, David Gesbert$^{(1)}$}
\IEEEauthorblockA{$^{(1)}$Communication Systems Department, EURECOM, Sophia Antipolis, France\\
$^{(2)}$Institute for the Wireless Internet of Things, Northeastern University, Boston, USA\\
Emails: \{omid.esrafilian, rakesh.mundlamuri, florian.kaltenberger, raymond.knopp, david.gesbert\}@eurecom.fr
}

\thanks{
$^{*}$ Equal contributions.
}
}

\maketitle

\begin{abstract} 
This paper considers the challenge of localizing ground users with the help of a radio-equipped unmanned aerial vehicle (UAV) that collects measurements from users. We utilize time-of-arrival (ToA) measurements estimated from the radio signals received from users collected by a UAV at different locations. Since the UAV's location might not be perfectly known, the problem becomes about simultaneously localizing the users and tracking the UAV's position. To solve this problem, we employed a least-squares simultaneous localization and mapping (SLAM) framework to fuse ToA data and the estimate of UAV location available from global positioning system (GPS). We verified the performance of the developed algorithm through real-world experimentation.

\end{abstract}


\section{Introduction}\label{sec:Intro}

In a wireless localization system, anchor nodes are used, which have precisely known positions. These nodes can either be stationary or mobile and collect various radio measurements from the radio frequency (RF) signals emitted by users in the network. The gathered measurements, including received signal strength (RSS), time of arrival (ToA), and angle of arrival (AoA), are employed for localization purposes \cite{zekavat2011handbook, mogyorosi2022positioning}.

Advancements in robotic technologies and the miniaturization of wireless equipment have led to the development of flying radio networks (FRANs). These networks utilize aerial base stations (BSs) or relays mounted on unmanned aerial vehicles (UAVs) to provide wireless connectivity to ground users \cite{gesbert2022uav,MozSaadBennNamDebb,mundlamuri2023integrated}. FRANs offer several advantages such as rapid deployment during emergencies or temporary crowded events, and the ability to provide connectivity in areas lacking network infrastructure. Unlike terrestrial radio access networks which use static BSs as anchor nodes, FRANs utilize UAV BSs as mobile anchor nodes. However, ensuring precise UAV location is essential, as the UAV location is not always accurately known and is subject to noise. Therefore, when using aerial mobile anchors, the challenge is not only localizing the users but also tracking the UAV location.

The use of UAV anchor nodes to collect radio measurements for localizing ground users has recently become popular\cite{ebrahimi2020autonomous,chen2020autonomous,jiang2020localization,esrafilian20203d, liang2022uav, le2020hybrid, SinYapISm, sinha2022impact, del2023first,del2023preliminary,del2022proof, esrafilian2023uavrssi, esrafilian2023uav}. One main advantage of using UAV anchors over static anchors is that UAVs, with their ability to move in three dimensions, can gather radio measurements in various geographic locations, thereby enhancing localization performance. In essence, UAVs in different locations can be viewed as virtual static anchors.

UAV-assisted user localization systems utilizing RSS measurements are studied in \cite{ebrahimi2020autonomous,chen2020autonomous,jiang2020localization,esrafilian20203d}. The authors in \cite{liang2022uav} assumed a multi-UAV-aided localization scenario in which a combination of ToA and AoA measurements are used for the localization of ground users. The UAVs trajectories are also optimized for further improvement of localization performance. Additionally, the authors in \cite{del2023first,del2023preliminary,del2022proof} investigate a multi-UAV-aided localization scenario in which a combination of time difference of arrival (TDoA) and global positioning system (GPS) measurements are used for the localization of ground users. A hybrid ToA and one-dimensional AoA localization approach, which requires elevation AoA estimations to integrate with ToA measurements, is proposed in \cite{le2020hybrid}. In \cite {SinYapISm, sinha2022impact}, the impact of the antenna radiation pattern on the communication channel between the UAV and ground users in a 3D localization system that utilizes time-based measurements is studied. 
In \cite{esrafilian2023uavrssi}, a UAV-aided user localization approach is introduced capitalizing on a mixture of RSS and ToA measurements. The radio channel is also assumed to be unknown, and the UAV trajectory is optimized to collect the most informative measurements to enhance localization performance. 
The authors in \cite{esrafilian2023uav}, employed a TDoA-based graph simultaneous localization and mapping (SLAM) approach to localize users utilizing the phase measurement collected by the UAV. 
In \cite{pfeiffer2023path}, the navigation of mobile robot to localize an mm-Wave wireless signal emitter is studied. The directionality properties of the signal is used to navigate the robot to localize the radio source node. 
The study in \cite{cheng2022ground} focuses on tracking the trajectory of a mobile user equipped with a single antenna. The location of the mobile user is treated as a virtual antenna, leading to the proposal of a single-antenna AoA estimation technique. This technique is designed to determine both the angle of arrival and the user's location over time. For the algorithm to function effectively, a rough estimate of the virtual antenna's location is necessary. However, if there is significant uncertainty in the initial estimate of the virtual antenna's location, the algorithm may fail.

In this paper, we propose a new algorithm for UAV-aided user localization systems. The location of the UAV is assumed not to be precisely known, hence the algorithm has to track the UAV location as well. To do so, we employed a least-squares SLAM framework, utilizing ToA and UAV location estimates available from GPS in a realistic 5G New Radio (NR) system. The ToA is estimated using the uplink sounding reference signal (SRS) in a 5G NR system with commercial user equipment (UE).

To the best of our knowledge, localization of users and tracking a mobile flying radio by exploiting Uplink ToA measurements has not been studied in the literature. Specifically, our contributions are as follows:
\begin{itemize}
\item An anchor-free localization system by utilizing a mobile flying radio node capable of collecting radio measurements from ground nodes is proposed.

\item A least-squares-based SLAM problem is formulated for fusing different types of measurements to jointly localize the users and track the UAV.  

\item The performance of the algorithm is verified via real-world implementation in a 5G NR open source system utilizing OpenAirInterface.
\end{itemize}

\begin{figure}[t]
\begin{centering}
\includegraphics[width=0.5\columnwidth]{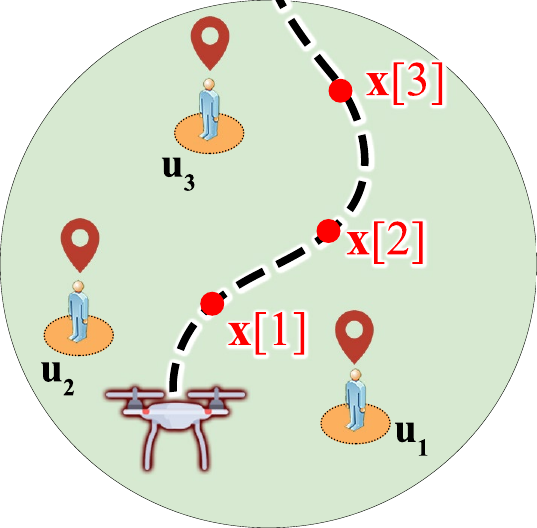}
\par\end{centering}
\caption{UAV-aided ground user localization system.\label{fig:SystemModel}}
\end{figure}

\section{System Model}\label{sec:SysModel}
We consider a scenario akin to the one shown in Fig. \ref{fig:SystemModel}, where a UAV equipped with radio devices can collect radio measurements from $K$ ground-level users in a given service area. The users are  
scattered over a given area and ${\bf u}_{k}=[x_{k},y_{k}]^{\Tr}\in\mathbb{R}^{2},\,k\in[1,K]$
denotes the $k$-th user's location. The users are assumed to be static and their locations are unknown.

The goal of the UAV is to localize the unknown users based on radio measurements collected over a mission of duration $T$. The UAV mission time is discretized into $N$ equal time steps. In the $n$-th time step, the UAV position is denoted by 
${\bf{x}}[n]=[x[n],y[n],z[n]]^{\Tr}\in\mathbb{R}^{3}$.
Moreover, the UAV location is obtained using a GPS receiver which is given by

\begin{equation}
    {\Hat{\bf{x}}}[n] = {\bf{x}}[n] + \eta,
\end{equation}
where $\eta $ is the GPS measurement noise and is modeled by a Gaussian random variable with $\mathcal{N}(0,\sigma_{gps}^2 {\bf{I}}_3)$ \cite{lee2016camera}, where ${\bf{I}}_{3}$ is the identity matrix of size $3 \times 3$.

\subsection{Channel Model}
This section describes the radio channel model between the UAV and ground users. 
Since air-to-ground channels are mainly exhibit few known dominant paths \cite{KhaGuvMao|_2019}, the channel between UAV at position ${\bf{x}}[n]$ and user location ${\bf{u}}_k$ can be modeled as 

\begin{equation}
g_{k}[n] = \sum_{l} a_{l, k}[n] \exp(-j \phi_{l, k}[n]) s_{k}\left[n - \tau_{l, k}[n]\right],
\label{eq:CH_Model}
\end{equation}
where $g_{k}[n], s_{k}[n]$ are, respectively, the received signal from user $k$ and the transmitted signal by the $k$-th user, $a_{l, k}[n]$ is the overall attenuation, $\phi_{l, k}[n] = 2 \pi f_c \tau_{l, k}[n]$ with $f_c$ equals to the carrier frequency, and $\tau_{l, k}[n]$ is the propagation delay of the $l$-th path between user $k$ and the UAV at time $n$. 

Assuming that the UAV and the users are stationary during a short period of time, then by taking several samples form $g_{k}[n]$, the parameters $a_{l, k}[n], \phi_{l, k}[n], \tau_{l, k}[n]$ can be estimated using classical channel estimation techniques. The estimated propagation delay can be modeled by Gaussian variables $ \Hat{\tau}_{l, k}[n] \sim \mathcal{N}(\tau_{l, k}[n],\sigma_{\tau}^{2})$, where $\sigma_{\tau}$, in general, is a function of the bandwidth and the signal-to-noise-ratio (SNR) of the signal between the UAV at time step $n$ and user $k$ \cite{esrafilian2023uav}. 

Note that, we assume the UAV can always establish a line-of-sight (LoS) connection to the users. We denote the LoS path with $l=0$. Defining $C$ as the speed of light, the true propagation delay for LoS path as a function of user and UAV locations is given by
\begin{equation}
    \tau_{0, k}[n] = \frac{\|{\bf{x}}[n] - {\bf{u}}_k \|}{C}. \label{eq:LoS_path_delay}
\end{equation}

\section{User Localization and UAV Tracking\label{sec:LearningLocalization}}

In this section, we present an algorithm for estimating user locations using the radio measurements gathered by the UAV. Let's represent the set of measurements collected by the UAV during the mission as $\mathcal{G} = \left\{\gamma_k[n], n\in [1,N],  k\in [1,K]\right\}$, where $\gamma_k[n]$ is a set of measurements collected by the UAV at time step $n$, defined as follows:
\begin{equation}
    \gamma_k[n] = \left({\Hat{\bf{x}}}[n], \Hat{\tau}_{0, k}[n] \right),
\end{equation}
where ${\hat{\mathbf{x}}}[n]$ represents the UAV location measured by the GPS and $\hat{\tau}_{0, k}[n]$ indicates the estimated propagation delay for the Line-of-sight (LoS) path between the UAV at time step $n$ and user $k$. It is important to note that the location of the UAV obtained by GPS may be prone to errors and might not be accurate in situations such as dense urban areas due to satellite signal obstruction by tall buildings. Therefore, the UAV location can also be estimated or improved along with the users locations. 

To this end, we use ToA measurements. However, the complexity of ToA-based localization methods increases linearly as the number of anchors grows (in our case, as more measurements are collected by the UAV). To address this issue, we retain the measurements if the distance between the UAV location when collecting those measurements is at least $\Delta$ meters apart.

Assuming that the collected measurements are independent and identically distributed (i.i.d) with respect to the channel and user positions, the negative log-likelihood of measurements results in 
\noindent 
\begin{equation}
\begin{aligned}\label{eq:localization_liklihood}
\mathcal{L} &=  \sum_{k=1}^{K}\sum_{n=1}^{N}
  \frac{1}{\sigma_{gps}^2}\left \| {\Hat{\bf{x}}}[n] - {\bf{x}}[n] \right\|^2 + \\
 & \sum_{k=1}^{K}\sum_{n=1}^{N} \frac{1}{\sigma_{\tau}^2}  \left | \Hat{\tau}_{0, k}[n] - \frac{\|{\bf{x}}[n] - {\bf{u}}_k \|}{C}\right|^2.
\end{aligned}
\end{equation}
The estimate of the unknown user and the UAV locations can then be obtained by solving
\begin{equation}\label{eq:SLAM_Opt_Org}
           \min_{\substack{{\bf{x}}[n],\,{{\bf{u}}}_k \\ \forall n, k}} \quad \mathcal{L}.
\end{equation}
\noindent Solving problem \eqref{eq:SLAM_Opt_Org} is challenging, since it is a simultaneous user localization and UAV tracking, and the objective function is non-linear and non-convex. To deal with this problem, we employ an iterative approach similar to the one presented in \cite{grisetti2010tutorial}, where at each iteration the problem first is locally linearized and then is solved. The algorithm then iterates until the convergence. In the following, we first introduce a general framework for solving optimization problems similar to \eqref{eq:SLAM_Opt_Org}, and then we will elaborate on how our problem can be solved with this framework.
Let's assume that we want to optimize the following problem
\begin{equation}\label{eq:Graph_SLAM}
           \min_{\vartheta} \quad \sum_i {\bf{e}}_i^{T}(\vartheta_i) {\bf{Q}}^{-1}_i {\bf{e}}_i(\vartheta_i).
\end{equation}
where $\vartheta = [\vartheta_0^{T}, \vartheta_1^{T}, \cdots]^{T}$ is a vector of all the unknown variables, ${\bf{e}}_i(\vartheta_i)$ is a vector function of the unknown variables $\vartheta_i$, and ${\bf{Q}}_i$ is a known diagonal matrix. By using the first-order Taylor approximation around an initial guess $\breve{\vartheta_i}$, we can write
\begin{equation} \label{eq:taylor_approx}
    {\bf{e}}(\breve{\vartheta_i} + \Delta \vartheta_i) \approx \breve{{\bf{e}}}_i + {\bf{J}}_i \Delta \vartheta_i 
\end{equation}
where $\breve{{\bf{e}}}_i \triangleq {\bf{e}}(\breve{\vartheta_i})$, and ${\bf{J}}_i$ is the Jacobian of ${\bf{e}}_i(\vartheta_i)$ computed in $\breve{\vartheta_i}$. By substituting \eqref{eq:taylor_approx} in \eqref{eq:Graph_SLAM}, we  have
\begin{equation}\label{eq:Graph_SLAM_approx}
           \min_{\vartheta} \quad \sum_i \breve{{\bf{e}}}_i ^{T} \breve{{\bf{e}}}_i + 2 \breve{{\bf{e}}}_i ^{T} {\bf{Q}}^{-1}_i {\bf{J}}_i \Delta \vartheta_i + \Delta \vartheta_i^{T} {\bf{J}}_i^{T} {\bf{Q}}^{-1}_i {\bf{J}}_i \Delta \vartheta_i.
\end{equation}
We can rewrite \eqref{eq:Graph_SLAM_approx} in a matrix form as follows
\begin{equation}\label{eq:Graph_SLAM_approx_mat}
           \min_{\vartheta} \quad \breve{{\bf{e}}} + 2 {\bf{b}}^{T} \Delta \vartheta + \Delta \vartheta ^{T}\, {\bf{H}}\, \Delta \vartheta,
\end{equation}
\noindent where $\breve{{\bf{e}}} \triangleq [\breve{{\bf{e}}}_0^{T}, \breve{{\bf{e}}}_1^{T}, \cdots]^{T}$,
${\bf{b}} = [\breve{{\bf{e}}}_0 ^{T} {\bf{Q}}^{-1}_0 {\bf{J}}_0, \breve{{\bf{e}}}_1 ^{T} {\bf{Q}}^{-1}_1 {\bf{J}}_1, \cdots]^{T}$, and ${\bf{H}}$ is a block diagonal matrix defined as
\begin{equation}\label{eq:graph_slam_hessian}
   {\bf{H}} \triangleq \text{diag}\left({\bf{J}}_0^{T} {\bf{Q}}^{-1}_0 {\bf{J}}_0, {\bf{J}}_1^{T} {\bf{Q}}^{-1}_1 {\bf{J}}_1, \cdots \right).
\end{equation}
Linear problem \eqref{eq:Graph_SLAM_approx} now can be solved and the solution is given by
\begin{equation}
    \vartheta^* = \breve{\vartheta} + \Delta \vartheta^* = \breve{\vartheta} - {\bf{H}}^{-1}\, {\bf{b}},
\end{equation}
where $\breve{\vartheta}=[\breve{\vartheta}_1^{T}, \breve{\vartheta}_2^{T}, \cdots]^{T}$ is a vector of initial guesses. This procedure will repeat until $\vartheta^*$ converges to local minima.
We now convert problem \eqref{eq:SLAM_Opt_Org} into a proper form for being solved with the above framework. To this end, we first define $\vartheta$ as follows
\begin{equation}
    \vartheta =\left [{\bf{x}}[1]^{T}, \cdots, {\bf{x}}[N]^{T}, {\bf{u}}_1^{T}, \cdots, {\bf{u}}_K^{T} \right]^T.
\end{equation}
\noindent We now reformulate problem \eqref{eq:localization_liklihood} as follows
\begin{equation}
\begin{aligned}
\mathcal{L} = \, & {\bf{e}}_{gps}^T  {\bf{Q}}_{gps}^{-1} {\bf{e}}_{gps} + {\bf{e}}_{\tau}^T  {\bf{Q}}_{\tau}^{-1} {\bf{e}}_{\tau},
\end{aligned}\label{eq:reformulated_to_graph_slam}
\end{equation}
\noindent where 
\begin{equation}
\begin{aligned}
{\bf{e}}_{gps} \triangleq & \left [\hat{{\bf{x}}}^{T}[1] - {\bf{x}}^{T}[1], \cdots, \hat{{\bf{x}}}^{T}[N] - {\bf{x}}^{T}[N]\right]^T, \\
{\bf{e}}_{\tau} \triangleq& \left [\hat{\tau_{0, 1}[1]} - \tau_{0, 1}[1], \cdots, \hat{\tau_{0, K}[N]} - \tau_{0, K}[N]\right]^T, \\
\end{aligned}
\end{equation}
\noindent and
\begin{equation}
\begin{aligned}
{\bf{Q}}_{gps} \triangleq & \, \sigma_{gps}^2 {\bf{I}}_N,
{\bf{Q}}_{\tau} \triangleq & \, \sigma_{\tau}^2 {\bf{I}}_{N},
\end{aligned}
\end{equation}
\noindent where ${\bf{I}}_{n}$ is the identity matrix of size $n \times n$. To solve \eqref{eq:reformulated_to_graph_slam} using the above framework, the GPS measurements are used for initializing the UAV location, and the users' locations are randomly initialized.
\section{System Design}
In this section, we elaborate on the equipment and tools used for designing our experimental platform.

\subsection{UAV Design}
To conduct the experiment, we have designed a custom drone capable of collecting measurements from users. This drone interacts seamlessly with an onboard 5G BS to gather ToA measurements, and it has the necessary flight time and payload capacity for our needs. It also communicates with a ground station application to receive control commands and report its status, including remaining battery life and location. The drone features a quad-rotor carbon body frame and is equipped with a Pixhawk 2 flight controller. For emergency situations, we have included a manual control option using a Futaba T8J radio controller.

\subsection{ToA Estimation using OpenAirInterface}

\begin{figure}[t]
\begin{centering}
\includegraphics[width=0.8\columnwidth]{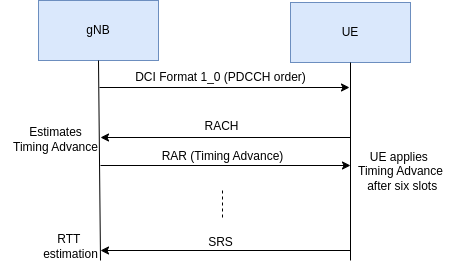}
\par\end{centering}
\caption{Signaling mechanism using PDCCH order\label{fig:pdcch_order}}
\end{figure}

The time of arrival is estimated at the 5G BS present on the UAV using OpenAirInterface\cite{KalaloAbhiLuh_20}. The time of arrival between the UAV position ${\bf{x}}[n]$ and the user location ${\bf{u}}_k$ is estimated from the round trip time (RTT). In the 5G BS, the RTT is estimated using a signaling mechanism described in Fig.~\ref{fig:pdcch_order}. 

In the signaling mechanism, the RTT estimation procedure between the UAV at a position ${\bf{x}}[n]$ and the user location ${\bf{u}}_k$ is initiated using a downlink control information (DCI) (ie., DCI Format 1\_0 for physical downlink control channel (PDCCH) order) and is estimated in two stages: 
\begin{enumerate}
    \item Coarse RTT estimation using random access channel (RACH).
    \item RTT refinement using SRS.
\end{enumerate}
In the coarse RTT estimation stage, the 5G BS present at the UAV can estimate a coarse RTT ${\tau_k^{c}[n]}$ from the timing advance ($\TA_k[n]$) value from RACH as follows,
\begin{equation}
 {\tau^c_k}[n] = \left(\TA_k[n]\times16\times64\times\text{T}_c\right)/ 2^{\mu},
\label{eq:RTTc}
\end{equation}
where, $\mu\in\{0,1,2,3,4,5\}$ is the numerology related to the subcarrier spacing $\Delta f=15.2^{\mu}\ \textrm{KHz}$, $\text{T}_c=\frac{1}{(\Delta f_{max}\times K_{max})}$, $\Delta f_{max}=480\ \text{KHz}$ is the maximum possible sub-carrier spacing and $K_{max}=4096$ is the maximum possible fast fourier transform (FFT) size in NR \cite{3gpp2018nr_38_211}. 

Further, in the second stage, a refined RTT estimate can be obtained from the estimated SRS channel impulse response ${\hat{g}_k}[n]$ that is received immediately after applying the timing advance as described in Fig.~\ref{fig:pdcch_order},
\begin{equation}
\tau_k^{r}[n]  = \frac{1}{f_s} \argmax \left|{\hat{g}_k}[n]\right|,
\label{eq:RTTr}
\end{equation}
where, ${\hat{g}_k}[n]$ is the estimated SRS channel frequency estimate between the UAV position ${\bf{x}}[n]$ at the user location ${\bf{u}}_k$ and $f_s$ is the sampling rate.

Finally, the RTT and the estimated propagation delay $\hat{\tau}_{0, k}[n]$ is obtained using \eqref{eq:RTTc} and \eqref{eq:RTTr} as follows,

\begin{equation}
    \text{RTT}_k[n] = {\tau^c_k}[n] + \tau^r_k[n],
\end{equation}
and
\begin{equation}
    \hat{\tau}_{0, k}[n] = \frac{\text{RTT}_k[n]}{2}.
\end{equation}

\begin{figure}[t]
\begin{centering}
\includegraphics[width=0.8\columnwidth]{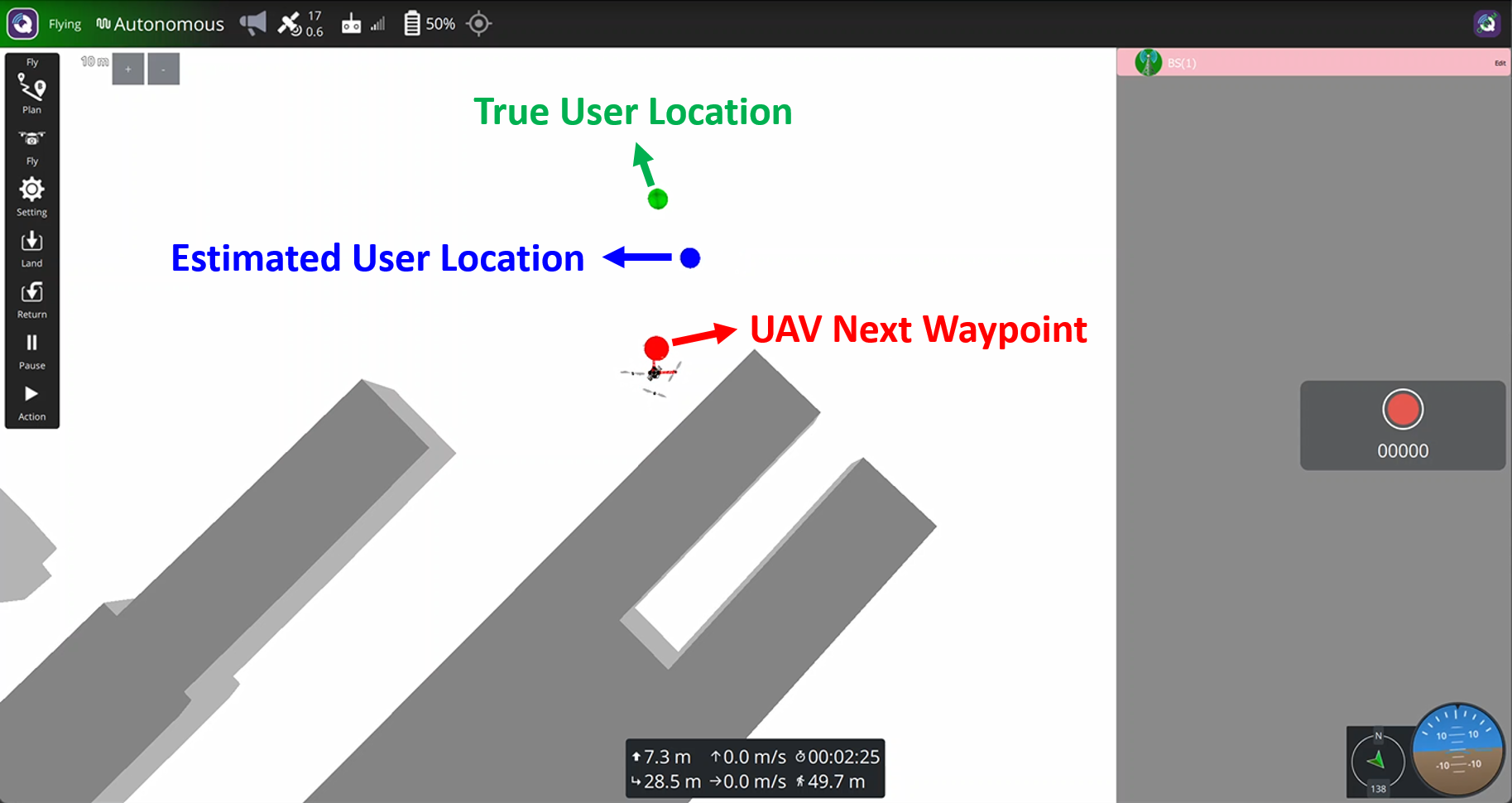}
\par\end{centering}
\caption{A screenshot of our connected robotics software platform, visualizing the 3D map, the current UAV and the estimated user locations, the UAV next way-point, and the true user location.\label{fig:QGC}}
\vspace{-4mm}
\end{figure}

\subsection{Connected Robotics Software Platform}
We have designed a connected robotics platform, an end-to-end 5G platform capable of controlling and monitoring a fleet of UAVs/robots over 5G links. This platform is based on an open-source project called QGroundControl (QGC) which is a widely used open-source project that provides a graphical user interface for monitoring, configuring, and controlling mobile robots like drones and rovers. Several new features such as a 3D viewer and network parameters monitoring, have been integrated into QGC, making it suitable for our connected-robotics use cases. The platform also hosts our entire localization algorithm. Once the UAV collects new measurements it is communicated to our platform on the ground via 5G links, and consequently, the UAV moves to a new way-point to collect more measurements. A screenshot of our software is shown in  Fig \ref{fig:QGC}.

\section{Experiment Results}\label{sec:results}
\begin{figure}[t]
\begin{centering}
\includegraphics[width=0.8\columnwidth]{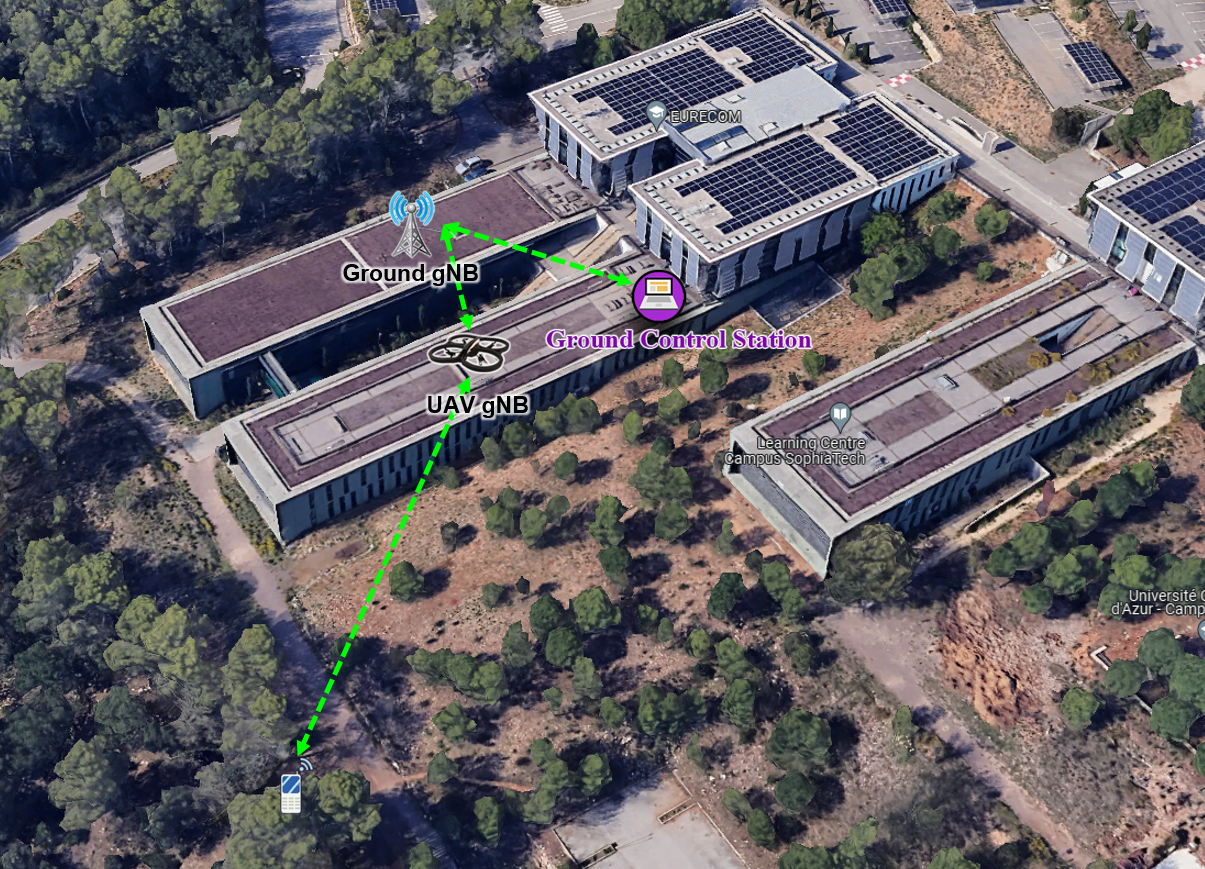}
\par\end{centering}
\caption{An example of our experiment setting where a single user is localized by a UAV 5G BS (a.k.a gNB).\label{fig:EXP_SHC}}
\vspace{-5mm}
\end{figure}

We conducted our experiments at the EURECOM premises to localize a single user on the ground. The user connects to the UAV 5G BS using a Quectel RM500Q-GL commercial 5G module. The UAV is equipped with an OAI-based 5G BS operating on band 41 with a 40 MHz bandwidth. To exchange measurements (such as ToA measurements and UAV position obtained by the onboard GPS) and control commands with our ground control station (i.e. connected robotics software platform), the UAV connects to a ground 5G BS operating in band 78 with a 30 MHz bandwidth. The ground control station can be connected to the ground 5G BS via cable or WiFi. The UAV collects data from the user approximately every $\Delta = 2$ meters. An example of the experiment setting is illustrated in Fig. \ref{fig:EXP_SHC}. Finally, regarding the GPS error, we set $\sigma_{gps} = 1$ m.

\begin{figure}[t]
\begin{centering}
\includegraphics[width=0.77\columnwidth]{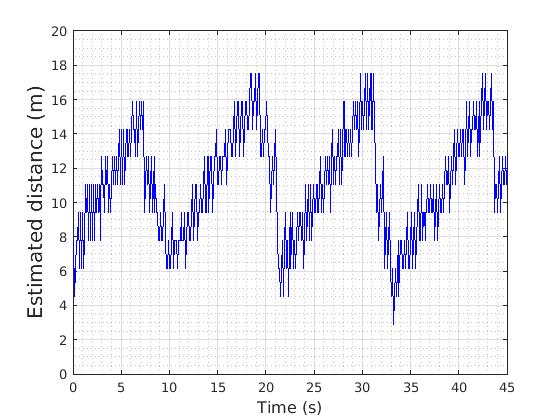}
\par\end{centering}
\caption{Effect of clock drift on ToA in commercial user.\label{fig:sawtooth}}
\vspace{-5mm}
\end{figure}

The estimated distance from the measured ToA over time using the signaling mechanism described in Fig.~\ref{fig:pdcch_order} and using a Quectel RM500Q-GL module can be seen in Fig.~\ref{fig:sawtooth}. It is worth noting that, a sawtooth behavior is observed in the estimated ToA over time while the user and the UAV remains stationary. In this sawtooth behavior, 
\begin{itemize}
    \item The rise in the ToA estimates stems from the clock drift between the UAV 5G BS and the user because the user does not correct its downlink (DL) timing immediately after receiving the DCI.
    \item The fall occurs when the user corrects its DL timing.
\end{itemize}
 This behavior in the commercial users using 3GPP Release-15 has an implementation-specific timing correction that corrects the DL timing only based on the conformance requirement\cite{3gpp2018nr_38_533} but not when the DCI is received. More details of the behavior and an improved signaling mechanism are addressed in \cite{mundlamuri2024novel}. However, utilizing the mechanism proposed in \cite{mundlamuri2024novel} is not possible with a commercial user, as used in this work, since it requires modifications to the source code of the 5G user module. In this work, we treat the error in estimating ToA due to this sawtooth behavior as noise which is captured in $\sigma_{\tau}$.

\begin{figure}[t]
\begin{centering}
\includegraphics[width=0.8\columnwidth]{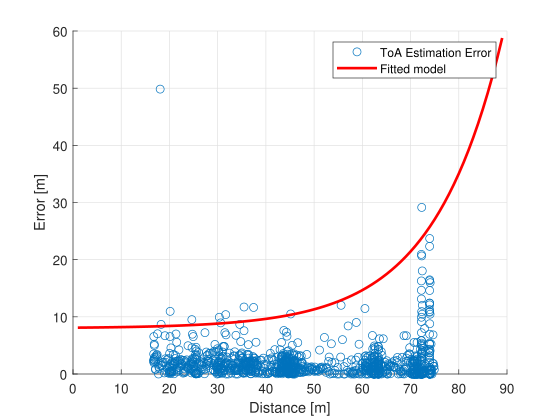}
\par\end{centering}
\caption{Absolute error of estimated distance between the UAV and the user based on measured ToA vs. true distance \label{fig:TOA_CoV}}
\vspace{-5mm}
\end{figure}

In Fig. \ref{fig:TOA_CoV}, we evaluate the absolute error of the estimated distance between the UAV 5G BS and the user based on measured ToA compared to the true distance. It is evident that the estimated distance experiences a sudden increase when the user moves outside the coverage range of the UAV 5G BS, which is attributed to the limited communication range. Therefore, we propose that an exponential function can effectively model this behavior, as illustrated by the solid line in Fig. \ref{fig:TOA_CoV}. Accordingly, this model can be used in \eqref{eq:localization_liklihood} to model the ToA estimate error $\sigma_{\tau}$.

\begin{figure*}
    \centering
    \begin{subfigure}[b]{0.25\textwidth}
        \includegraphics[width=\textwidth]{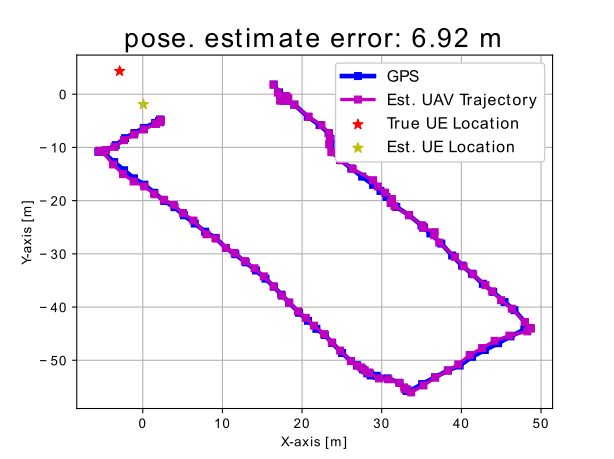}
        \caption{Top View}
        \label{fig:real_pos}
    \end{subfigure}
    \begin{subfigure}[b]{0.25\textwidth}
        \includegraphics[width=\textwidth]{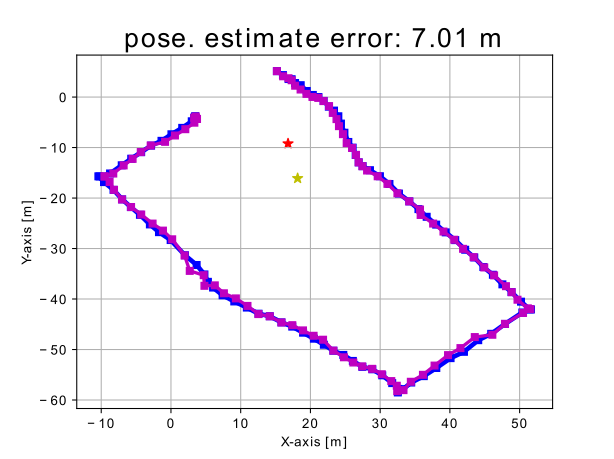}
        \caption{Top View}
        \label{fig:pred_pos_siam}
    \end{subfigure}
        \begin{subfigure}[b]{0.25\textwidth}
        \includegraphics[width=\textwidth]{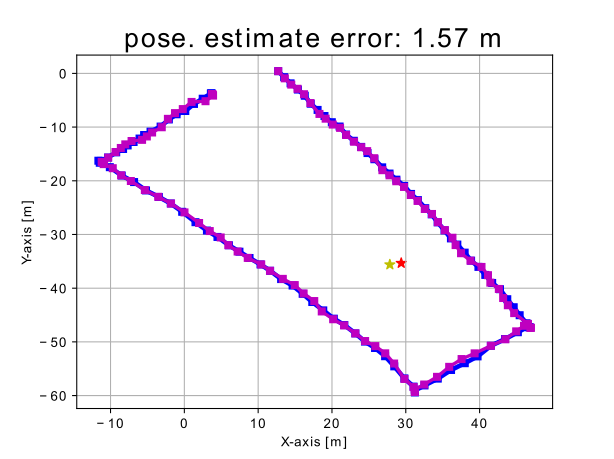}
        \caption{Top View}
        \label{fig:real_pos}
    \end{subfigure}
    \begin{subfigure}[b]{0.25\textwidth}
        \includegraphics[width=\textwidth]{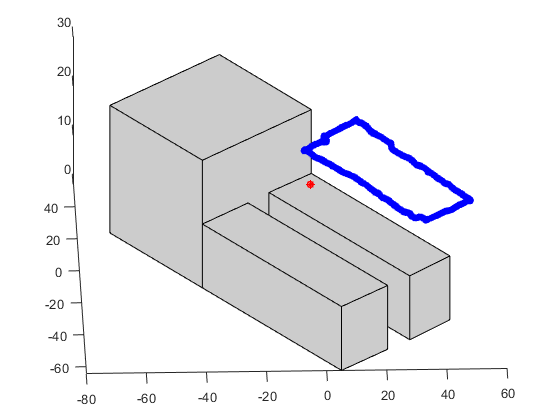}
        \caption{Side View}
        \label{fig:pred_pos_siam}
    \end{subfigure}
        \begin{subfigure}[b]{0.25\textwidth}
        \includegraphics[width=\textwidth]{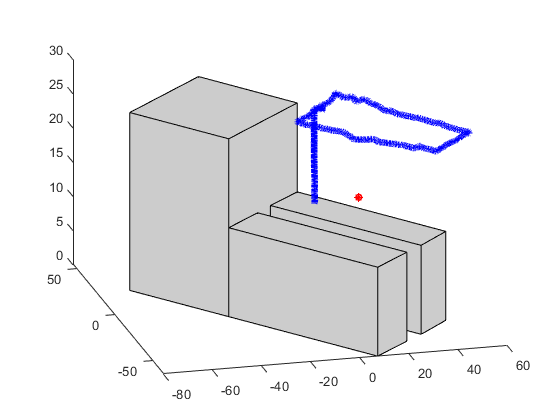}
        \caption{Side View}
        \label{fig:real_pos}
    \end{subfigure}
    \begin{subfigure}[b]{0.25\textwidth}
        \includegraphics[width=\textwidth]{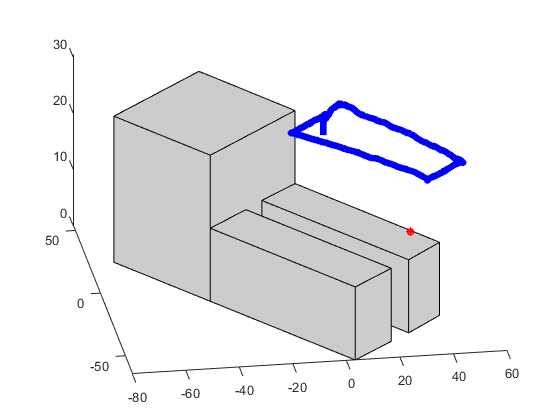}
        \caption{Side View}
        \label{fig:pred_pos_siam}
    \end{subfigure}
    \caption{Figures (a) to (c) are the top view of the taken trajectory, while figures (d) to (f) show the side view of the scenarios for each experiment. The user position (pose.) estimates absolute error for each experiment is reported in the title of each figure.}
    \label{fig:los_cresults}
    \vspace{-4mm}
\end{figure*}

In Fig. \ref{fig:los_cresults}, we have demonstrated various scenarios with different user locations for the joint localization of the user and the UAV when the user is always LoS to the UAV. We can observe that the user can be localized within the expected accuracy based on the signal bandwidth. Additionally, the estimated UAV trajectory closely aligns with the position measured by the GPS due to the fact that the UAV at all times has a clear line of sight connection to the satellites, and consequently, the GPS uncertainty remains low compared to ToA measurements $\sigma_{gps} \ll \sigma_{\tau}$. 

\begin{figure}
    \centering
    \begin{subfigure}[b]{0.24\textwidth}
        \includegraphics[width=\textwidth]{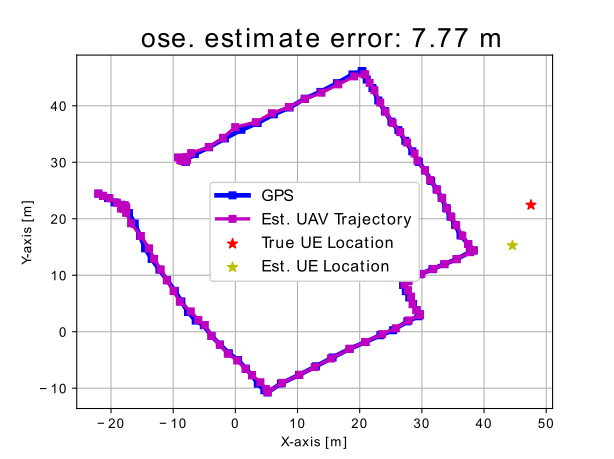}
        \caption{Top View}
        \label{fig:real_pos}
    \end{subfigure}
    \begin{subfigure}[b]{0.24\textwidth}
        \includegraphics[width=\textwidth]{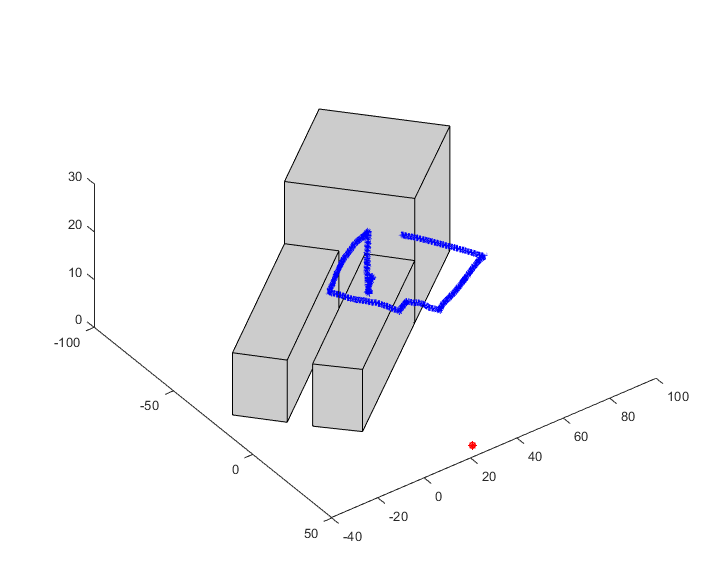}
        \caption{Side View}
        \label{fig:pred_pos_siam}
    \end{subfigure}
    \caption{The top view and the side View of the scenario where the UAV is NLoS to the user at some locations.}
    \label{fig:nlos_results}
    \vspace{-4mm}
\end{figure}

In Fig. \ref{fig:nlos_results}, we present the results from a scenario where the connection between the user and the UAV is obstructed by buildings in certain areas. Our findings demonstrate that, even when the measurements include Non-LoS (NLoS) data, the algorithm is capable of compensating for these NLoS measurements. As a result, it still provides an accurate estimate of the user's location, similar to the results obtained in completely LoS scenarios.


\section{Conclusions}\label{sec:conclusion}
This paper addressed the challenge of localizing ground users using a radio-equipped UAV gathering measurements from users. We focus on ToA measurements derived from the radio signals the UAV receives at various locations. Since the precise location of the UAV may not be fully known, the task involves simultaneously localizing the users and tracking the UAV's position. To tackle this issue, we employed a least-squares SLAM framework that integrates ToA data with the UAV's location estimate obtained from GPS. We validated the performance of the proposed algorithm through real-world experiments.

In our future work, we will investigate the problem of synchronization error caused by clock drift, which leads to a sawtooth pattern in ToA estimation.

\section*{Acknowledgment}
\footnotesize This work is partially funded by 
the 5G-OPERA project through the German Federal Ministry of Economic Affairs and Climate Action (BMWK) as well as the French government as part of the ``France 2030" investment program.

\bibliographystyle{IEEEtran}
\bibliography{literature.bib}

\end{document}